\documentclass[reprint,amsmath,amssymb,aps,superscriptaddress]{revtex4-2}

\usepackage{amssymb,amsthm,bm,xcolor}
\usepackage{graphicx}
\usepackage{dcolumn}
\usepackage{bm}
\usepackage{url}
\usepackage{hyperref}
\usepackage{physics}
\usepackage{float}
\usepackage{times}
\usepackage[normalem]{ulem}
\usepackage{braket}
\usepackage{booktabs}
\usepackage{siunitx}
\usepackage{soul}

\hypersetup{
	colorlinks=true,
	urlcolor=purple,
	citecolor=purple,
	linkcolor=purple,
	breaklinks}
    
\setstcolor{red}

\newcommand{\Renyi}{R\'{e}nyi}

\begin{document}

\preprint{APS/123-QED}

\newcommand{\zju}{School of Physics, ZJU-Hangzhou Global Scientific and Technological Innovation Center, \\and Zhejiang Key Laboratory of Micro-nano Quantum Chips and Quantum Control, Zhejiang University, Hangzhou, China}
\newcommand{\hefei}{Hefei National Laboratory, Hefei, China}
\newcommand{\hkust}{The Hong Kong University of Science and Technology (Guangzhou), Nansha, Guangzhou, China}

\author{Yiren Zou} 
\thanks{These authors contributed equally.}
\affiliation{\zju}

\author{Hong-Kuan Xia}
\thanks{These authors contributed equally.}
\affiliation{\hkust}

\author{Aosai Zhang} 
\thanks{These authors contributed equally.}
\affiliation{\zju}

\author{Xuhao Zhu}
\affiliation{\zju}
\author{Feitong Jin}
\affiliation{\zju}
\author{Qingyuan Wang}
\affiliation{\hkust}
\author{Yu Gao}
\affiliation{\zju}
\author{Chuanyu Zhang}
\affiliation{\zju}
\author{Ning Wang}
\affiliation{\zju}
\author{Zhengyi Cui}
\affiliation{\zju}
\author{Fanhao Shen}
\affiliation{\zju}
\author{Zehang Bao}
\affiliation{\zju}
\author{Zitian Zhu}
\affiliation{\zju}
\author{Jiarun Zhong}
\affiliation{\zju}
\author{Gongyu Liu}
\affiliation{\zju}
\author{Jia-Nan Yang}
\affiliation{\zju}
\author{Yihang Han}
\affiliation{\zju}
\author{Yiyang He}
\affiliation{\zju}
\author{Jiayuan Shen}
\affiliation{\zju}
\author{Han Wang}
\affiliation{\zju}
\author{Yanzhe Wang}
\affiliation{\zju}
\author{Jiahua Huang}
\affiliation{\zju}
\author{Xinrong Zhang}
\affiliation{\zju}
\author{Sailang Zhou}
\affiliation{\zju}

\author{Hang Dong}
\affiliation{\zju}
\author{Jinfeng Deng}
\affiliation{\zju}
\author{Yaozu Wu}
\affiliation{\zju}
\author{Zixuan Song}
\affiliation{\zju}

\author{Hekang Li}
\affiliation{\zju}
\affiliation{\hefei}
\author{Zhen Wang}
\affiliation{\zju}
\affiliation{\hefei}
\author{Chao Song}
\affiliation{\zju}
\affiliation{\hefei}
\author{Qiujiang Guo}
\affiliation{\zju}
\affiliation{\hefei}

\author{Pengfei Zhang}
\email{pfzhang@zju.edu.cn}
\affiliation{\zju}
\affiliation{\hefei}

\author{Guo-Yi Zhu}
\email{guoyizhu@hkust-gz.edu.cn}
\affiliation{\hkust}

\author{H. Wang}
\email{hhwang@zju.edu.cn}
\affiliation{\zju}
\affiliation{\hefei}


\title{Teleportation transition of surface codes on a superconducting quantum processor}

\begin{abstract}
   The topological surface code is a leading candidate for harnessing long-range entanglement to protect logical quantum information against errors, and teleportation of logical states is desirable for robust quantum information processing. Nevertheless, scaling up the surface code in quantum teleportation poses a formidable challenge to experiment. 
   Here on a superconducting quantum processor with 125 qubits, we demonstrate the robust teleportation of topological rotated surface code prepared by a linear-depth unitary circuit, with code distances up to $7$. 
   We obtain the teleportation phase diagram by tuning the local entangling gates uniformly across a finite threshold.
   Furthermore, we show that the entangling threshold can be boosted by coherent qubit rotations that inject magic resources beyond the Clifford regime, restoring the duality symmetry of the topological phase, which serves as a guiding principle to minimize the entanglement resource. 
   Our results shed light on simulating and leveraging topological quantum matter on quantum devices, and pave the way to the ultimate goal of distributed fault tolerant quantum computation. 
\end{abstract}

\maketitle


Quantum error correction is a promising route toward fault-tolerant quantum computation. 
A leading candidate is the surface code~\cite{kitaev2003fault,bravyi1998quantum,Krinner2022Nature,Acharya2025Nature,He2025Phys.Rev.Lett.,Marques2022Nat.Phys.}, which encodes a logical qubit into a long-range entangled~\cite{chen2010local} state of many physical qubits. It leverages topological order~\cite{wen19choreographed,Song2018Phys.Rev.Lett.,Satzinger2021Science} to robustly protect quantum information against errors~\cite{Preskill2002}, where errors are detected by measuring two types of stabilizers called syndromes, and are corrected by corresponding Pauli operations. Scaling up the surface code is challenging but necessary to achieve higher tolerance against the errors. 
The scaling also poses a challenge for the many-body generalization of the quantum teleportation~\cite{Bennett1993}, which is an exotic quantum phenomenon that harnesses entanglement resource to enable quantum communication between two systems.
A potential elementary module of the distributed quantum computing architecture in the future requires the teleportation of a topological surface code, carrying its logical qubit~\cite{Ryan-Anderson2024Science}. 
However, in the presence of incoherent noise and coherent errors, experimentally demonstrating the many-body teleportation remains an open question. 

\begin{figure*}[!htbp]
\centering
\includegraphics[width=\linewidth]{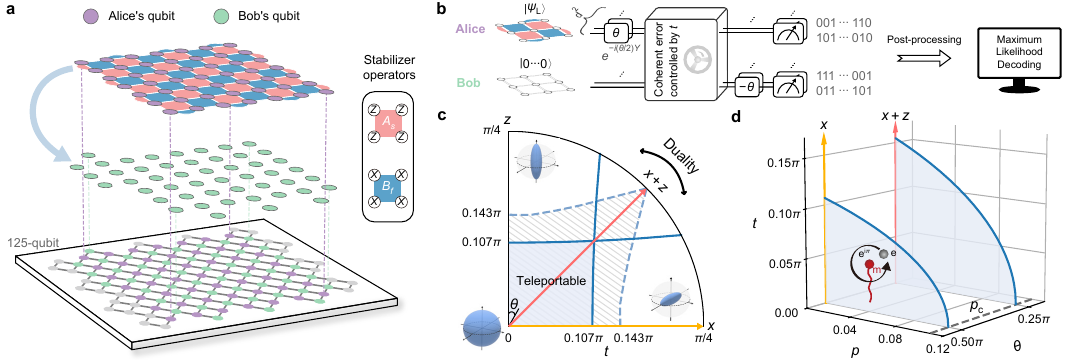}
\caption{
{\bf Experimental setup and conceptual illustration. }
{\bf a}, Our 125-qubit quantum processor and two embedded distance-7 surface codes, labelled Alice (purple) and Bob (green). Each surface code comprises 49 qubits, with the logical subspace defined by $Z$-type stabilizers and $X$-type stabilizers. In the subsequent experiments, logical states are teleported from Alice to Bob, as indicated by the light blue arrow. Alice and Bob are physically interlaced to facilitate this state teleportation.
{\bf b}, Schematic illustration of the tunable teleportation protocol. The process begins with the initialization of Alice in an encoded logical state, while Bob's qubits are prepared in the physical product state $\ket{0\cdots0}$. A parameterized coupling operation, defined by parameters $\theta$ and $t$, is then applied between the systems to induce entanglement. The measurement outcome of Alice is then communicated to Bob, upon which local corrections are applied. The final logical information is retrieved via a classical decoding algorithm.
{\bf c}, Topological phase diagram under coherent errors in the $XZ$-plane. When the injected coherent error magnitude $t$ is small (shaded blue region), logical information can be faithfully recovered after imperfect teleportation, demonstrating that the topological order of the surface code is preserved. Rotating the coherent error axis from the $X$ direction (orange arrow) to the $X+Z$ direction (red arrow), significantly enhances the entangling threshold. The solid lines indicate the entangling threshold for teleportation, while the dashed line indicates the learning threshold of Alice. 
{\bf d}, Phase diagram under incoherent bit-flip noise. In addition to coherent errors, we analyse the performance of logical teleportation in the presence of incoherent noise, modelled here by an independent bit-flip channel acting on the physical qubits. The teleportation of topological logical state becomes infeasible when the noise strength surpasses the critical threshold $p_\mathrm{c}=0.109$.
}
\label{fig:teleportation_conception}
\end{figure*}

Here, we experimentally demonstrate a teleportation protocol for a logical qubit of the surface code with imperfect entanglement resource. 
To capture the essence of the practical teleportation over long distances~\cite{Pan2017} where entanglement resource is imperfect, we tune the entangling gates continuously to inject coherent errors, which enables charting out the many-body teleportation transition across the {\it entangling threshold}~\cite{Eckstein2024PRXQuantum}. 
Further, with insights from the $\mathbb{Z}_2$ lattice gauge theory~\cite{FradkinShenker79,kitaev2003fault} and topological order~\cite{chen2010local}, we show {\it experimentally} how one can enhance the entangling threshold by restoring the electric–magnetic duality~\cite{Eckstein2024PRXQuantum,Wang25selfdual}. 
In practice, the teleportation experiment is implemented by utilizing a high-performance 125-qubit superconducting quantum processor (Fig.~\ref{fig:teleportation_conception}a). This processor features tunable nearest-neighbour couplings~\cite{Yan2018Phys.Rev.Appl.}, allowing for rapid and high-fidelity two-qubit gates. The median fidelities for single-qubit and two-qubit gates are $99.94\%$ and $99.64\%$, respectively, which are essential for both the preparation of surface code logical states and their subsequent teleportation. Additionally, we achieve a median readout fidelity of $99.08\%$ by pumping each qubit from the state $ \ket{1} $ to the state $ \ket{2} $ prior to the readout pulse.

Our setup consists of two groups of qubits, referred to as Alice and Bob, each arranged in a $ d \times d $ square lattice~\footnote{For $d=7$, there is slight rearrangement of Bob's qubits on the boundary, see Supplemental Information for details.}, where $ d $ represents the code distance. 
We first prepare a surface code on Alice prior to the teleportation, while Bob remains in the product state $\ket{0 \cdots 0}$. 
A surface code~\cite{bravyi1998quantum} can be characterized by a set of stabilizer operators, which are divided into two distinct categories (see Fig.~\ref{fig:teleportation_conception}a). 
A $Z$-type stabilizer $ A_s = \prod_{j \in s} Z_{j} $ is the product of Pauli $ Z $ operators acting on the qubits surrounding a red face. 
An $X$-type stabilizer $ B_f = \prod_{j \in f} X_{j} $ is the product of Pauli $X$ operators for the qubits around a blue face. 
These stabilizers constrain the Hilbert space formed by $d^2$ qubits to a two-dimensional logical subspace.
Specifically, a logical state, represented as $ \ket{\psi_\mathrm{L}} $, is defined as a simultaneous $+1$ eigenstate of all stabilizers: $ A_s \ket{\psi_\mathrm{L}} = B_f \ket{\psi_\mathrm{L}} = +1 \ket{\psi_\mathrm{L}} $.
The logical subspace is defined by the logical $Z_\mathrm{L}$ and $X_\mathrm{L}$ operators, which commute with all stabilizers but anti-commute with each other.
An arbitrary logical state can be generated using a circuit whose depth increases linearly with the code distance~\cite{Satzinger2021Science,Zhang2025arxiv}. For further details on the state preparation circuits, please refer to the Supplementary Information (SI) Section~S4.

The teleportation protocol involving three parties~\cite{Bennett1993} can be simplified into a two-party state transfer protocol, as illustrated in Fig.~\ref{fig:teleportation_conception}b, where the surface code on Alice is transferred to Bob as a whole. 
In contrast to previous experiments on logical state teleportation~\cite{Ryan-Anderson2024Science,Lacroix2025Nature}, in which the target qubits are initially encoded into a logical qubit, we prepare Bob's qubits here in the simple product state $\ket{0 \cdots 0}$, such that the logical qubit is teleported accompanying the collective topological order.
The interlacing between Alice and Bob qubits is designed for convenience of transversally and uniformly tuning the entanglement between Alice and Bob, while minimizing unnecessary errors, in demonstrating the teleportation transition. Our aim is to chart out the landscape for the required minimal entanglement resource for such many-body teleportation~\cite{Eckstein2024PRXQuantum} (see Fig.~\ref{fig:teleportation_conception}c).
The presence of incoherent noise, however, reduces the entangling threshold (Fig.~\ref{fig:teleportation_conception}d). 
Under a bit-flip noise channel, described by $\mathcal{N}_X\left(\rho\right) = \left(1-p\right) \rho + p\, X \rho X$, the topological phase is completely destroyed when the error rate $p$ exceeds a critical value ($p_\mathrm{c}=0.109$~\cite{Preskill2002, Wan25nishimori}~\footnote{The asymptotic thresholds for coherent and incoherent errors in teleportation experiments, obtained using simulations involving one million qubits, are $p_\mathrm{c}=0.1092212(4)$, $t_\mathrm{c}\left(\theta=\pi/2\right) =0.1072128(2)\pi$, and $t_\mathrm{c}\left(\theta=\pi/4\right) =0.1548014(3)\pi$.}), making faithful quantum teleportation impossible even with perfect entanglement resource.
Consequently, a fundamental prerequisite for a successful demonstration of teleportation is the fidelity of generated logical states (see Methods for a nonlinear benchmark of the prepared noisy surface codes for modest code distances $d = 2, 3$).

\begin{figure*}[t!]
   \centering
   \includegraphics[width=\linewidth]{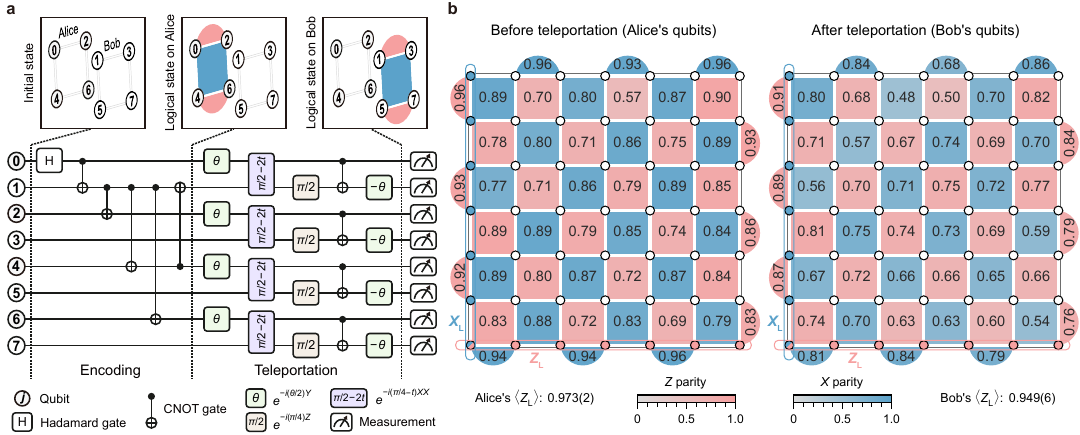}
   \caption{
   {\bf Circuit and stabilizer parities before and after teleportation.}
   {\bf a}, Schematic circuit for state preparation and teleportation for code distance $2$. 
   Initializing from a product state, after the first stage
   Alice's qubits (white circles with even numbers) are prepared in the logical state $\ket{0_\mathrm{L}}$ of a surface code, while Bob's qubits (white circles with odd numbers) stay in the product state $\ket{0\cdots0}$. 
   In the second stage, the unitary gates entangle Alice's and Bob's qubits one-on-one (i.e., {\it transversally}). After measuring out the qubits of Alice, the state of surface code is successfully transferred from Alice to Bob i.e., teleportation when involving only two parties. 
   The gate parameter $t$ plays the role of ``coherent error", that tunes the entangling, between Alice and Bob, from maximal ($t=0$) to minimal ($t=\pi/4$). Here the feedback correction of the measurement outcomes are performed in advanced by the CNOT gates before the measurement of Alice's qubits, such that after these CNOT layers, Alice and Bob are effectively disentangled when $t=0$, as shown in the schematic. 
   The rotation angle $\theta$ is a knob that does not play a role when the entangling is maximal $t=0$, but it can tune the threshold of coherent error, as will be shown later. 
   {\bf b}, Measured $X$-type and $Z$-type stabilizer parity values for the logical state $\ket{0_\mathrm{L}}$ using Alice's qubits (left) and Bob's qubits (right). To calculate the $X$ and $Z$ parities, we simultaneously measure all of Alice's (or Bob's) qubits in the $X$ and $Z$ basis, respectively, over $4\times10^4$ measurement shots, followed by readout error mitigation (SI Section~S1C). 
   Mean parity: $0.82(8)$ for Alice and $0.68(8)$ for Bob. Using the ML decoder, we estimate the expectation value of the logical operator $Z_\mathrm{L}$ to be $0.973(2)$ for Alice and $0.949(6)$ for Bob. The corresponding logical error rate is given by $\left(1 - \left\langle Z_\mathrm{L} \right\rangle\right)/2$.
   }
   \label{fig:fig_parity_and_circuit_2}
\end{figure*}

\section*{Teleportation transition}

\subsubsection*{Tunable teleportation protocol}

We now sweep the coherent error strength in our protocol to validate the existence of a stable and long-range phase. 
To do that, we use the circuit depicted in Fig.~\ref{fig:fig_parity_and_circuit_2}a to teleport the logical state generated on Alice's side to Bob's side. 
First, a unitary gate $U\left(\theta, \phi\right)$ is transversally applied to each physical qubit of Alice. The unitary gate rotates the Pauli $Z$ to an arbitrary direction on the Bloch sphere, $\hat{\sigma}^{\theta, \phi} = \sin(\theta) \cos(\phi)X + \sin(\theta)\sin(\phi)Y + \cos(\theta)Z$. 
In the following experiments, we focus on the case where $\phi=0$, confining the direction on the $XZ$ plane.
Next, a two-qubit \emph{entangling} gate $R_{XX}\left(\pi/2 - 2t\right)$ parameterized by $t$ is used to tune the microscopic entangling between Alice and Bob: $t=0$ corresponds to $R_{XX}\left(\pi/2\right)$ that maximally entangles Alice's and Bob's qubits transversally, and increasing $t$ reduces the entangling, where $t=\pi/4$ reaches the trivial limit without any entangling. One can view $t$ as a source of coherent error that weakens the entanglement resource between Alice and Bob for teleportation.
In the standard teleportation or state transfer protocol~\cite{Bennett1993}, each qubit in Bob is manipulated according to the measurement outcome of the corresponding qubit in Alice. 
As the feedback operation is yet to be developed for our hardware system, here we choose to postpone Alice's measurement and use a layer of controlled-NOT (CNOT) gates to replace the classical communication.
The final $R_Y(-\theta)$ gates rotate Bob's qubits back to the original direction for measurement.

Conditional upon the bitwise $Z$-expectation values ($s_j^Z=1$ or $-1$) in Alice at the output, the input state of Alice is transferred to Bob with a deformation factor of $\exp\left({\beta} \sum_j s^Z_j \hat{\sigma}_j^{\theta,\phi}/{2}\right)$, where $\tanh(\beta) = \sin(2t)$ for $t\in[0,\pi/4]$~\cite{Eckstein2024PRXQuantum}~\footnote{Note that the regime of $t>\pi/4$ is equivalent up to an extra layer of deterministic Pauli gates, which is not considered here.}.
The coherent error $t$ that decreases the entangling between Alice and Bob is thus mapped to a weak measurement error for Bob's received state. 
If $t=\beta=0$, the topological properties of the state in Alice are maximally teleported to Bob's side, while Alice obtains featureless random measurement outcomes. In contrast, for $t=\pi/4$, $\beta=\infty$, the teleportation is blocked. Nevertheless, Alice effectively measures the qubits of the surface code in $\hat{\sigma}^{\theta,\phi}$ basis, and then {\it classical} outcomes are cloned to Bob.

\subsubsection*{Teleportation below the threshold}

We first shut off the coherent error at $t=0$ to investigate the performance of our protocol in the presence of practical noise.
The largest two surface codes we can achieve on this device have a code distance of $d=7$, which require a total of 98 data qubits. We begin by examining the parities of the logical state $ \ket{0_\mathrm{L}} $ generated on Alice (Fig.~\ref{fig:fig_parity_and_circuit_2}b, left). Here, the $Z$-type and $X$-type parities are defined as the expectation values of the $A_s$ and $B_f$ stabilizer operators, respectively. The measured parities are similar to those reported in recent experiments~\cite{Satzinger2021Science,Xu2023Chin.Phys.Lett.}, and are obtained after readout error mitigation. 
Under a phenomenological uniform bit-flip noise model, the expected parity values scale as $\left(1 - 2p\right)^w $ with $p$ denoting the error rate and $w \in \{2, 4\}$ corresponding to the stabilizer weight, which explains the observed difference between the parity values in the bulk and on the boundaries. 
Consequently, we square the weight-2 stabilizer values before calculating the average, and obtain a mean parity value of $0.82(8)$. After state teleportation, the measured $Z$ and $X$ parities on Bob's qubits are shown in the right panel of Fig.~\ref{fig:fig_parity_and_circuit_2}b, with a mean value of $0.68(8)$. 
The measured bitstrings are processed by a tensor network (TN)-based maximum likelihood (ML) decoder to generate an error-corrected output. 
The logical error rate is measured to be $0.014(1)$ for Alice and $0.025(3)$ for Bob.  
The slight increase in the logical error rate can be attributed to the additional errors introduced by the teleportation circuit.
The high parity values and associated low logical error rates indicate that the teleportation is operating below the threshold.

\subsubsection*{Engineering the threshold}

\begin{figure}[tbp]
\centering
\includegraphics[width=3.5 in]{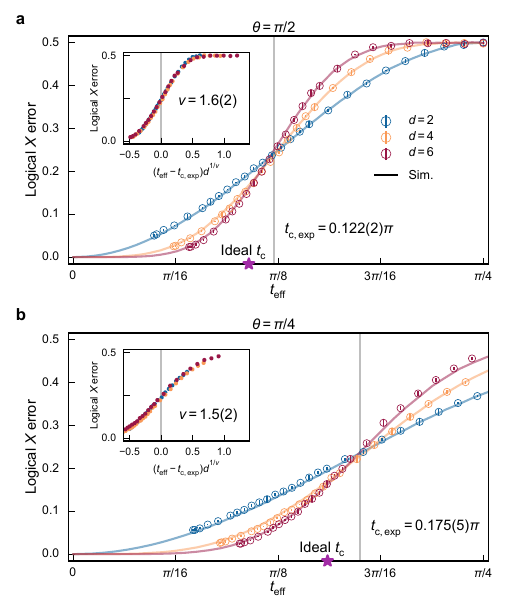}
\caption{
{\bf Teleportation transition by tuning entanglement.} 
{\bf a}, Logical $X$ error rate as a function of the effective entangling parameter $t_\mathrm{eff}$ for errors injected along the $X$ axis ($\theta=\pi/2$). 
The parameter $t_\mathrm{eff}$ combines both coherent and incoherent errors (see the main text).
Experimental results (circles with error bars) fit to a pseudo-threshold of $0.122(2)\pi$ (vertical grey solid line) that is slightly higher than the asymptotic threshold of $0.107\pi$ (purple star).
The pseudo-threshold is extracted from a finite-size scaling analysis shown in the inset. Solid colored lines: noiseless simulation results.
{\bf b}, Logical $X$ error rate as a function of the effective entangling parameter $t_\mathrm{eff}$ for errors injected along the $X+Z$ axis ($\theta=\pi/4$). At this angle, the system exhibits an electric-magnetic duality, leading to a significant enhancement of the entangling threshold. The experimentally observed pseudo-threshold is $0.175(5)\pi$, while the asymptotic threshold is $0.155\pi$. 
}
\label{fig:fig_logical_error_MLD_2me}
\end{figure}

We first consider the injection of coherent errors along the $X$ axis ($\theta=\pi/2$) of the phase diagram in Fig.~\ref{fig:teleportation_conception}c. 
In general, $t$ not only drives a transition of teleportation to Bob, but also a learning transition of Alice, determining whether her measurement records enable learning of the logical state~\cite{Zhu25learncode}. Alice's learnability transition is shown by dashed line in Fig.~\ref{fig:teleportation_conception}. This can be captured by the coherent information~\cite{Nielsen96coherentinfo, Lloyd97coherentinfo}, i.e., entropy of the mixed logical state conditioned upon her measurement outcome. We show this in SI Section~S3C, which is consistent with Nishimori threshold $t_\mathrm{c}=0.143\pi$~\cite{Eckstein2024PRXQuantum}. On the other hand, Bob's received state can be directly measured and decoded.
When Alice's measurement record is discarded, the weak measurement operator is twirled into an effective Pauli channel $\mathcal{N}(\rho) = \cos^2(t) \rho + \sin^2(t) X \rho X$. 
If without additional noise, it shows an entangling threshold at $t_\mathrm{c}=0.107\pi$~\cite{Eckstein2024PRXQuantum}, corresponding to the critical noise probability of Nishimori transition $p_\mathrm{c}=0.109$. The gap between Alice's learning threshold and Bob's decoding threshold can in principle be closed by designing a certain correlated decoder, which we leave for future investigations.

In the experiments, two types of errors are present: (1) a coherent error introduced via the tunable teleportation scheme, controlled by the parameter $t$ of the two-qubit $R_{XX}$ gate, and (2) an accumulated incoherent error, modelled as uniform bit-flip noise with an error rate of $p$. To encapsulate both contributions, and since we mainly tune the coherent error, we define an effective entangling parameter, $t_\mathrm{eff}$, satisfying $\sin^2(t_\mathrm{eff}) = \sin^2(t) + p\cos(2t)$ as detailed in the SI Section~S3. The term proportional to $p$ drives $t_\mathrm{eff}$ towards $\pi/4$, leading to a residual error even when $t=0$. Check that $p=1/2$ results in the same consequence as $t=\pi/4$ here. 
In Fig.~\ref{fig:fig_logical_error_MLD_2me}a, we mainly show the logical error rates as a function of $t_\mathrm{eff}$ for code distances $d = 2, 4, 6$, while leaving the data for $d=3, 5, 7$ in SI Section~S3C. 
The experimental data agrees closely with noiseless numerical simulations (solid lines in Fig.~\ref{fig:fig_logical_error_MLD_2me}a), while the residual effective entangling parameter increases with the code distance, due to the noise arising from increasing circuit depths.
Using a finite-size scaling analysis, we extract a threshold of $t_\mathrm{c,\, exp}\left(\theta=\pi/2\right) = 0.122(2)\pi$, which slightly deviates from the ideal value of $0.107\pi$~\cite{Wan25nishimori} in the thermodynamic limit.
We attribute this overestimation of the threshold to the fact that there are more $Z$-type stabilizer syndrome information for even code distance. In comparison, the odd code distances underestimate the threshold, see SI Section~S3C. 
Alternative decoding results, obtained using both the TN-based Bayesian decoder and the minimum-weight perfect matching decoder, are presented in SI Section~S5.

The teleportation threshold is mathematically equivalent to the \emph{phase transition} of the underlying quantum state of matter. A practical method to enhance the protection of the logical state against errors is therefore to foster competition between electric and magnetic excitations, leveraging the anyon statistics emergent from the topological order of the surface code~\cite{FradkinShenker79, Zhu19, Eckstein2024PRXQuantum, Grover24selfdual}.
By rotating the axis of the coherent error from the $X$ direction toward the $Z$ direction, one can effectively transform the induced excitations from purely magnetic to purely electric~\cite{kitaev2003fault, Zhu19}. As the fraction of magnetic excitations gradually decreases, the logical $Z$ fidelity becomes increasingly resilient to coherent errors, leading to a larger and eventually an infinite error threshold (Fig.~\ref{fig:teleportation_conception}c). 
Meanwhile, the fraction of electric excitations correspondingly increases, making the logical $X$ fidelity less robust, and therefore the coherent error can destroy the underlying topological order of the surface code. 
Notably, when the error axis is along the $X+Z$ direction, the system exhibits an electric-magnetic self-duality~\cite{FradkinShenker79, Zhu19}, characterized by invariance under Hadamard transformations. 
In the context of quantum computation, such rotation injects \emph{magic}~\cite{Kitaev05magic, niroula2024magic} resource to the physical circuit beyond the Clifford class~\cite{Gottesman97}, which is optimal at angle $\theta=\pi/4$. 
At this optimal angle, the weak measurement operator is twirled into a dephasing channel in non-Clifford basis:
\begin{equation}
   \mathcal{N}(\rho) = \cos^2(t)\rho +\sin^2(t)\frac{X+Z}{\sqrt{2}}\rho \frac{X+Z}{\sqrt{2}} \ ,
\end{equation}
by discarding Alice's outcomes. 
That is, the effective Pauli bit-flip error contributed by the weak measurement is reduced by a factor of 2: $\sin^2(t)/2$ and the effective entangling parameter $t_\mathrm{eff}$ is modulated to be $\sin^2(t_\mathrm{eff}) = \sin^2(t) +p\left(1+\cos\left(2t\right)\right)$.
At this self-dual point, the threshold is significantly increased to $t_\mathrm{c,\, exp}\left(\theta=\pi/4\right) =0.175(5)\pi$, as shown in Fig.~\ref{fig:fig_logical_error_MLD_2me}b, in comparison with the known ideal value of $0.155\pi$ for Nishimori threshold.
However, while the coherent information in Ref.~\cite{Eckstein2024PRXQuantum} shows the ultimate high threshold $0.250\pi$, here the decoding threshold is smaller because we use only one type of syndromes.

\section*{Teleporting logical magic states}

\begin{figure}[tbp]
\centering
\includegraphics[width=\columnwidth]{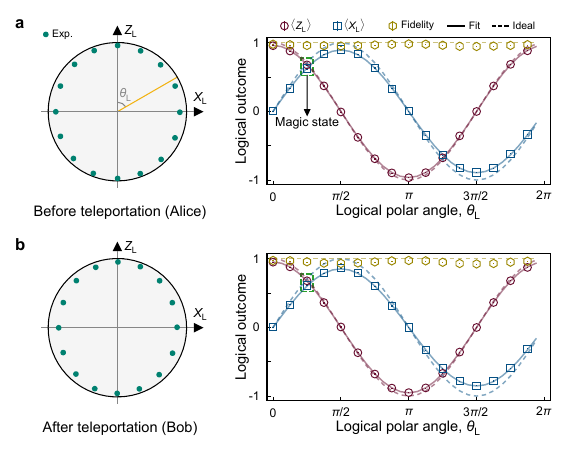}
\caption{
{\bf Teleportation of arbitrary logical states.} 
{\bf a}, Logical measurement results before teleportation (Alice, $d = 3$). These states are chosen to span the $XZ$-plane of the logical Bloch sphere, parameterized by the logical polar angle $\theta_{\rm{L}}$ (left panel). The right panel shows the measured expectation values of the logical operators $X_{\rm{L}}$ and $Z_{\rm{L}}$ for these states. Solid lines are fitting results, and dashed lines indicate noiseless ideal results. The $H$-type magic state at $\theta_{\rm{L}}=\pi/4$ is highlighted by the green dashed box. 
Average fidelity: $0.96(1)$.
{\bf b}, Logical measurement results after teleportation (Bob, $d = 3$). 
The average state fidelity of Bob's logical qubit is $0.95(2)$, only slightly lower than that of the initial states on Alice, reflecting a high-fidelity teleportation process. In all plots, error bars are significantly smaller than the corresponding marker size.}
\label{fig:arbitrary_logical_state}
\end{figure}

The teleportation protocol is expected to work for arbitrary unknown logical quantum states. 
A key example is the teleportation of a logical {\it magic} state~\cite{Kitaev05magic}, which is a crucial resource for achieving universal fault-tolerant quantum computation. In our experiments, we generate a family of states lying on the $XZ$-plane of the logical Bloch sphere by setting the azimuthal angle $\phi_{\mathrm{L}}=0$ and varying the polar angle $\theta_{\mathrm{L}}$, such that the states are expressed as $\ket{\psi_\mathrm{L}}=\cos\left(\theta_\mathrm{L}/2\right)\ket{0_\mathrm{L}} + \sin\left(\theta_\mathrm{L}/2\right)\ket{1_\mathrm{L}}$.
This family includes, in particular, the $H$-type magic state corresponding to $\theta_{\mathrm{L}} = \pi/4$. 
These states are experimentally prepared by a non-fault-tolerant state injection method~\cite{Satzinger2021Science}, whose fidelity can in principle be further improved by magic state distillation~\cite{Li2015NewJ.Phys.,Ye2023Phys.Rev.Lett.,SalesRodriguez2025Nature}. 

We validate these states by measuring the expectation values of the logical operators $X_{\rm{L}}$ and $Z_{\rm{L}}$ (Fig.~\ref{fig:arbitrary_logical_state}a), with ML decoder.
The experimental results are found to be captured by the functions $\langle X_{\mathrm{L}} \rangle = \left(1 - 2p_{Z}\right) \sin (\theta_{\mathrm{L}})$ and $\langle Z_{\mathrm{L}} \rangle = (1 - 2p_{X}) \cos (\theta_{\mathrm{L}})$, which fit two asymmetric {\it logical} error probabilities: $p_{X}=0.019(1)$ and $p_{Z}=0.053(1)$.
This error asymmetry is further confirmed using the $H$-type magic state, where the ideal expectation values of $\langle X_{\mathrm{L}} \rangle$ and $\langle Z_{\mathrm{L}} \rangle$ are equal. Instead, our experiment yields a lower $\langle X_{\mathrm{L}} \rangle=0.62(1)$ compared to $\langle Z_{\mathrm{L}} \rangle=0.68(1)$. 
Numerical simulation corroborates that this asymmetry should be attributed to the deep circuit of unitary-based state preparation; notably, the asymmetry is eliminated when the logical state is prepared by the (stabilizer) measurement-based approach instead of the unitary approach (SI Section~S4C).
The logical state fidelity $\mathcal{F}_{\rm{L}}$, given by 
\begin{equation}
\mathcal{F}_{\mathrm{L}} = \frac{1}{2}\left(1 + \cos \theta_{\mathrm{L}} \langle Z_{\mathrm{L}} \rangle + \sin \theta_{\rm{L}} \langle X_{\mathrm{L}} \rangle \right),
\end{equation}
is also shown in Fig.~\ref{fig:arbitrary_logical_state}a.
We then teleport these states from Alice to Bob using parameters $\theta=\pi/2$ and $t=0$, with the results shown in Fig.~\ref{fig:arbitrary_logical_state}b. A similar analysis on the teleported states gives slightly higher error probabilities of $p_{X}=0.024(1)$ and $p_{Z}=0.074(1)$, which indicates that the teleportation process adds noise while preserving the error asymmetry.

\section*{Discussion and outlook}

These experiments demonstrate the key ingredients required for teleporting an arbitrary unknown logical state of a surface code, even with imperfect entanglement resource. By introducing a tunable teleportation knob, our approach charts out the phase diagram to determine the essentially required entanglement, in the presence of noise. Notably, the entangling threshold is found to be significantly enhanced when the weak measurement error axis is rotated from the $X$ (or $Z$) axis to the $X+Z$ axis, a clear manifestation of electric-magnetic duality.

Looking ahead, several important and intriguing avenues remain to be explored. 
One promising direction is to incorporate a fast \emph{mid-circuit} measurement protocol~\cite{Jeffrey2014Phys.Rev.Lett.,Walter2017Phys.Rev.Appl.,Spring2025PRXQuantum} in conjunction with an efficient qubit reset technique~\cite{Reed2010Appl.Phys.Lett.,Egger2018Phys.Rev.Appl.,McEwen2021Nat.Commun.,Marques2023Phys.Rev.Lett.,Lacroix2025Phys.Rev.Lett.}. 
This would reduce the circuit depth required to prepare the topological ordered state from a linear depth down to a constant depth. Besides, it would enable both Alice and Bob to perform adaptive circuits to suppress incoherent noise before and after the teleportation process, also allowing classical communication between Alice and Bob. 
By initializing the system in a cleaner state and extracting additional syndrome information, we anticipate further improvement in overall process fidelity, as well as a more favorable entangling threshold.
Another compelling frontier is the extension of this protocol to a truly nonlocal setting, namely, to perform teleportation across a long physical distance~\cite{Bouwmeester1997, Hanson2014, Pan2017} between two superconducting processors or between two codes separated on a large-scale processor. It is expected that as distance is increased, the entangling resource is weakened, corresponding to increasing $t$ in our phase diagram.
Successfully demonstrating fault-tolerant logical state transfer between {\it distant} modules would be a critical step toward building scalable quantum computing networks.

In conclusion, our work provides a comprehensive experimental characterization of logical state teleportation and establishes a foundational toolkit for probing the resilience of encoded information against tailored noise. These results and techniques not only advance our understanding of topological quantum matter but also open new possibilities for realizing robust distributed quantum computing.

\section*{Methods}

\subsubsection*{Device performance}

\begin{figure}[!htbp]
\centering
\includegraphics[width=3.5 in]{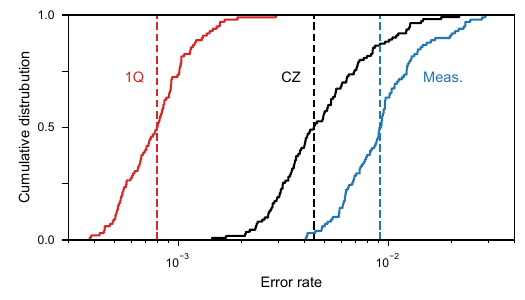}
\caption{
{\bf Cumulative distributions of the operation errors for the distance-7 teleportation experiment}. Red: Pauli errors for simultaneous single-qubit gates. Black: Pauli errors for simultaneous CZ gates. Blue: identification errors for state discriminations. 
The vertical dashed lines indicate median fidelities: $99.94\%$ for single-qubit gates, $99.64\%$ for two-qubit gates, and $99.08\%$ for readout. 
}
\label{fig:figS_teleportation_operate_error}
\end{figure}

Our experiments are performed on a flip-chip superconducting quantum processor. As shown in Fig.~\ref{fig:teleportation_conception}a, $98$ qubits are used to implement the distance-$7$ teleportation experiment, with the measured energy relaxation time $T_1$ and spin-echo dephasing time $T_2^\mathrm{SE}$ having median values of about \SI{70.1}{\micro\second} and \SI{18.2}{\micro\second}, respectively. The native universal gateset consists of arbitrary single-qubit rotations and two-qubit controlled-$\pi$ phase (CZ) gate. The operations in our experimental circuits, including $R_{XX}$ and CNOT gates, are decomposed into this native gateset.
Single-qubit rotations are implemented via a \SI{20}{\nano\second} $XY$ microwave drive combined with virtual $Z$ rotations. For two-qubit operations, the couplers between adjacent qubits are dynamically tuned to realize fast CZ gates within \SI{32}{\nano\second}. 
Finally, the states of all qubits are read out simultaneously using individual readout resonators that are capacitively coupled to each qubit. Figure~\ref{fig:figS_teleportation_operate_error} shows the gate and measurement errors of our distance-$7$ teleportation experiment.

\subsubsection*{Relative entropy of logical states}

\begin{figure}[!htpb]
   \centering
   \includegraphics[width=\columnwidth]{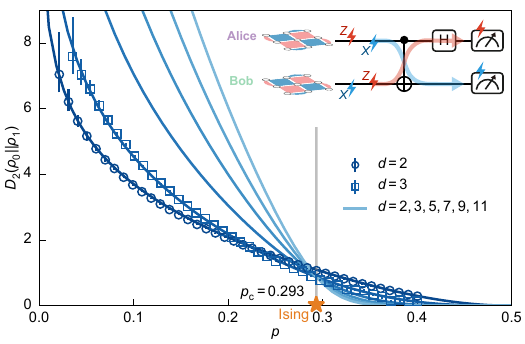}
   \caption{
      {\bf {\Renyi} relative entropy between orthogonal logical states, for code distances $d=2$ and $3$.} Circles and squares indicate the experimental data from Bell measurement, where the data with higher noise probability $p$ is obtained with classical error injection. Solid lines show the numerical result for larger sizes. 
      The orange star indicates Ising critical point at $p_\mathrm{c} =1 / \left(2+\sqrt{2}\right) \approx 0.293 $.
      The inset schematic shows the circuit preparing Alice and Bob into logical states $\left(\rho_0, \rho_0\right)$ or $\left(\rho_0, \rho_1\right)$ and subsequently performing Bell measurements, and how mid-circuit Pauli errors propagate to affect the readout. 
}
\label{fig:relative_entropy}
\end{figure}

Here, we show a nonlinear benchmark of the prepared noisy surface codes for modest code distances $d=2,3$ on our device. 
Namely, we evaluate the relative {\Renyi} entropy between two logical states supported on two sets of qubits following Alice and Bob partitioning. It is given by
\begin{equation}
D_2(\rho_0||\rho_1) = -\ln \text{tr}(\rho_0 \rho_1) + \ln \text{tr} (\rho_0^2) \ ,
\label{eq:D2}
\end{equation}
where $\rho_0$ and $\rho_1$ are the density matrices corresponding to the prepared logical states $\ket{0_\mathrm{L}}$ and $\ket{1_\mathrm{L}}$, respectively.
The purity-like quantities, $\text{tr}(\rho_0^2)$ and $\text{tr}(\rho_0\rho_1)$, can be measured by the standard joint Bell measurements for each pair of qubits (see SI Section~S2 for details). 
The measured values of $\text{tr}\left(\rho_0^2\right)$ are $0.846(2)$ for $d=2$ and $0.528(1)$ for $d=3$, which fit to effective noise strengths of $p=0.0211(4)$ for $d=2$ and $p=0.0351(1)$ for $d=3$, shown by the markers of smallest $p$ in Fig.~\ref{fig:relative_entropy}.
Furthermore, we manually inject error to the experimental bitstrings in the post-processing, obtaining $D_2$ versus stronger noise. 

The experimental data agrees well with numerical simulation, with uniformly injected independent bit-flip and phase-flip noise, $\mathcal{N}\left(\rho\right) = \mathcal{N}_X\left(\rho\right) \circ \mathcal{N}_Z\left(\rho\right)$, where $ \mathcal{N}_X\left(\rho\right) = \left(1-q\right) \rho + q X \rho X$ and likewise for $Z$. 
They can be propagated to the final measurement resulting in effective readout error of $p=2q(1-q)$, because of the entangling gate for Bell measurement. 
The finite size scaling behaviour of the numerical simulation and the experimental data indicate a sharp phase transition. Indeed, Eq.~\eqref{eq:D2} can be mapped (see SI Section~S2) to a classical 2D Ising model, which is exactly solved to yield an Ising critical point at $q_\mathrm{c}=0.178$, or equivalently readout error strength $p_\mathrm{c} = 0.293$.

\vspace{10pt}

\noindent {\bf Acknowledgments} The device was fabricated at the Micro-Nano Fabrication Center of Zhejiang University. 
The electronics for controlling the superconducting quantum processors were developed by Hangzhou Logical Qubit Technology. 
We acknowledge the support from the Quantum Science and Technology-National Science and Technology Major Project (grant no.~2021ZD0300200), 
the National Natural Science Foundation of China (grant nos.~92365301, 12404574, 12274368, 12274367, 12322414, and 12404570), 
the Zhejiang Leading Goose (Lingyan) Project (grant nos.~2026C02A2004),
and the Zhejiang Pioneer (Jianbing) Project (grant no.~2025C01019). 
H. Wang is also supported by the New Cornerstone Science Foundation through the XPLORER PRIZE.
G.Z. would like to thank Simon Trebst for collaboration on related projects and for helpful discussions. 
G.Z., H.X. and Q.W. acknowledge the support of NSFC-Young Scientists Fund (grant no.~12504181) and Start-up Fund of HKUST(GZ) (grant no.~G0101000221), also support in part by grant no.~NSF PHY-2309135 to the Kavli Institute for Theoretical Physics (KITP). 

\vspace{10pt}

\noindent {\bf Author contributions} G.Z., H.X. and Q.W. proposed the ideas and conducted the theoretical and numerical analysis and TN-based ML decoder for the experimental bitstrings; Y.Z. and A.Z. carried out the experiments and analysed the experimental data under the supervision of P.Z. and H.~Wang; H.L. fabricated the device supervised by H.~Wang; G.Z., P.Z., Y.Z., H.X., A.Z., and H.~Wang co-wrote the manuscript; H.~Wang, P.Z, Z.W., C.S., Q.G., Y.Z., A.Z., X.~Zhu, F.J., Y.G., C.Z., N.W., Z.C., F.S., Z.B., Z.Z., J.Z., J.-N.Y., Y.~Han, Y.~He, G.L., J.S., Han~Wang, Y.~Wang, J.H., X.~Zhang, S.Z., H.D., J.D., Y.~Wu, and Z.S. contributed to the experimental setup. All authors contributed to the discussions of the results.

\bibliography{reference}

\end{document}


\clearpage
\setcounter{table}{0}
\renewcommand{\thetable}{S\arabic{table}}%
\renewcommand{\thefigure}{S$\the\numexpr\value{figure}$}%
\setcounter{equation}{0}
\renewcommand{\theequation}{S\arabic{equation}}%
\setcounter{section}{0}
\renewcommand{\thesection}{S\arabic{section}}%
\setcounter{page}{1}
\begin{center}
    \textbf{\large Supplementary Information for \\
    ``Teleportation transition of surface codes on a superconducting quantum processor''
    }\\[.2cm]
\end{center}

\tableofcontents

\section{Experimental setup and noise characterization}

\subsection{Effective incoherent noise strength}
\label{sec:effective_noise}

We estimate the effective incoherent noise strength $p$ for our teleportation experiment using the measured parities without applying readout error correction. Since the stabilizer expectation values depend on the stabilizer weight, we extract $p$ via
\begin{equation}
    1 - 2p = \left(\dfrac{\sum_{s \in \mathcal{S}_2} \langle A_s \rangle^2 + \sum_{s \in \mathcal{S}_4} \langle A_s \rangle}{\#\,\mathcal{S}_2 + \#\,\mathcal{S}_4}\right)^{1/4},
    \label{eq:incoherent_noise_strength}
\end{equation}
where $\mathcal{S}_w$ denote the set of $Z$-type faces defined by $w \in \{2, 4\}$ qubits and $\#\,\mathcal{S}_w$ is the number of elements in $\mathcal{S}_w$. The extracted values are provided in Table~\ref{tab:incoherent_noise_strength}.

\begin{table*}[!htbp]
    \setlength{\tabcolsep}{6pt}
    \renewcommand{\arraystretch}{1.5}
    \centering
    \caption{\label{tab:incoherent_noise_strength} 
    {\bf Effective incoherent noise strength for each code distance.}
    These noise strength values are derived from measured stabilizer parities with teleportation parameters fixed at $\theta = \pi/2$ and $t=0$.
    The values being a few percent are due to the joint action of the mid-circuit noises and the readout errors that lead to the infidelity of the measurement outcomes. 
    }
    
    \begin{tabular}{c|cccccc}
        \hline \hline 
        Code distance, $d$ & 2 & 3 & 4 & 5 & 6 & 7 \\
        \hline
        Effective incoherent noise strength, $p$ & 0.0237(5) & 0.0306(5) & 0.0364(4) & 0.0380(3) & 0.0484(4) & 0.0505(2) \\
        \hline \hline
    \end{tabular}
\end{table*}

The unitary circuit of preparing the surface code scales linearly with the code distance. Assuming that the accumulated noise increases with the circuit depth, the finite noise threshold poses an upper bound on the circuit depth and thus the code distance. 
We can approximate the effective noise strength as a linear function $p(d) = a(d+d_0)$, where the maximum code distance that remains below the threshold $p_\mathrm{c}=0.109$ is then given by 
\begin{equation}
    d_\mathrm{max} = \left\lfloor\frac{p_\mathrm{c}}{a} - d_0\right\rfloor.
\end{equation}
Fitting the experimentally observed noise levels in our quantum processor provides $a = 0.0054$ and $d_0 = 2.5$. Consequently, the maximum distance for a surface code below the threshold is approximately $d_\mathrm{max} = 17$, which can be further increased by improving qubit coherence times and gate fidelities.

\begin{figure}[!htbp]
\centering
\includegraphics[width=3.5 in]{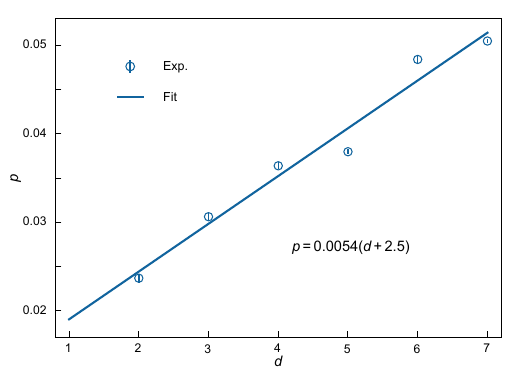}
\caption{
\textbf{Effective bit-flip error strength accumulated during circuit implementation.} 
The effective error strength $p$ is modelled as a linear function of the code distance $d$, yielding the relationship $p=0.0054 (d + 2.5)$. Experimental results are represented by circles with error bars, while the dashed line indicates the linear fit.
}
\label{fig:figS_p_vs_d}
\end{figure}

\subsection{Qubits allocation}
\label{sec:space}

\begin{figure}[!htbp]
\centering
\includegraphics[width=3.5 in]{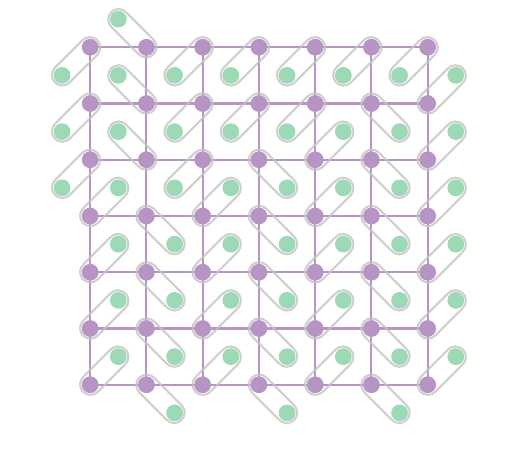}
\caption{\textbf{Spatial arrangement of Alice’s and Bob’s qubits for the distance-7 teleportation experiment.}
Purple circles denote Alice’s qubits, and green circles denote Bob’s qubits. Grey outlines indicate the one-to-one pairing of corresponding qubits used in the teleportation experiment.
}
\label{fig:figS_teleportation_group}
\end{figure}

To implement the teleportation experiment, we interlace Alice's and Bob's qubits such that each of Bob's qubits is positioned at the immediate bottom-right lattice site relative to its corresponding qubit in Alice. However, due to the presence of non-functional qubits along the boundaries, this precise configuration is feasible only for code distances $d \leq 6$. Consequently, as illustrated in Fig.~\ref{fig:figS_teleportation_group}, we introduce a spatial rearrangement of Bob's qubits, positioning each at a nearest-neighbour lattice site (not precisely the bottom-right) relative to its corresponding qubit in Alice.

\subsection{Readout error mitigation}
\label{sec:readout_error_mitigation}

Here we show details for readout error mitigation for measuring the stabilizer expectation values to validate our prepared surface codes targetting the stabilizer states, which should be distinguished from the task of decoding a logical qubit. 
Measurement of superconducting qubits is vulnerable to various types of errors.
As a result, the measured observables can be significantly degraded with respect to their actual values. The dominant error mechanisms include: (1) separation errors arising from a finite signal-to-noise ratio in the amplifying chain, which limit the state distinguishability, (2) qubit energy relaxation occurring during the readout window, which can change the qubit state before the measurement is completed, and (3) unwanted coherent transitions between qubit states, induced, for instance, by stray couplings or measurement-induced leakage out of the computational subspace. 

In quantum error correction circuits, measurement errors are typically addressed through decoding algorithms that analyse the measured syndrome patterns to identify and correct the errors. However, this approach is not directly applicable to measurement outcomes of the physical qubits themselves. For quantities such as single-qubit expectation values and multi-qubit parities, they are not topologically protected and there is no associated decoding algorithm. The measurement results must be corrected by an explicit readout-error mitigation procedure.

The effect of readout errors can be quantitatively modelled by a response matrix~\cite{Bravyi2021Phys.Rev.A}. In this description, the noisy measurement process is represented as a classical channel that maps the actual bitstring $s'$ to the observed bitstring $s$ with a certain conditional probability. Denoting by $P_\mathrm{a}(s')$ the actual probability of obtaining bitstring $s'$  and by $P_\mathrm{m}(s)$ the measured probability of $s$, the two distributions are related via
\begin{equation}
    P_\mathrm{m}(s) 
    = \sum_{s^\prime} M(s \mid s^\prime)\, P_\mathrm{a}(s^\prime),
\end{equation}
where $M(s \mid s^\prime)$ are the elements of the response matrix $\mathbf{M}$. By construction, $M(s \mid s^\prime)$ specifies the probability of measuring the measurement outputs $s$ given that the system is prepared in the state $\ket{s'}$. In matrix notation, this relationship is written as $\mathbf{P}_\mathrm{m} = \mathbf{M}\,\mathbf{P}_\mathrm{a}$.

In the experiment, the response matrix is calibrated by preparing the system in a complete set of product states $\{\ket{s'}\}$ and performing many repeated measurements for each preparation. For a given input state $\ket{s'}$, the observed empirical frequencies of the outcomes $s$ define the column of $\mathbf{M}$ associated with $s'$. Assuming that the response matrix is invertible and that the calibration is sufficiently accurate, one can then estimate the actual distribution by applying the inverse matrix to the measured data $\mathbf{P}_\mathrm{a} = \mathbf{M}^{-1}\,\mathbf{P}_\mathrm{m}$. Subsequently, corrected expectation values of observables, such as multi-qubit parities, can be obtained by evaluating them with respect to the reconstructed distribution $\mathbf{P}_\mathrm{a}$.

While conceptually straightforward, this full response matrix approach faces severe scalability limitations. The dimension of the response matrix is $2^N \times 2^N$, so both its experimental calibration and numerical inversion require resources that grow exponentially with the qubit number $N$. Consequently, a brute-force characterization and inversion of the full response matrix is feasible only for small systems and quickly becomes impractical as the number of qubits increases.

In superconducting qubit platforms, however, the dominant contribution to measurement errors arises from processes that affect each qubit approximately independently, such as individual readout misclassification or single-qubit relaxation during measurement. Under this assumption, the measurement noise can be accurately described by a tensor-product noise model, in which the full response matrix is constructed as
\begin{equation}
    \mathbf{M} = \bigotimes_{n=1}^{N} \mathbf{M}_n,
\end{equation}
where $\mathbf{M}_n$ is the single-qubit response matrix associated with qubit $Q_n$. This tensor product structure dramatically reduces the experimental and computational overhead. Each single-qubit response matrix is a $2 \times 2$ matrix that can be calibrated independently using only the two single-qubit preparations $\{\ket{0}, \ket{1}\}$, and the inversion of $\mathbf{M}$ reduces to inverting $N$ single-qubit response matrices and taking their tensor product, $\mathbf{M}^{-1} = \bigotimes_{n=1}^{N} \mathbf{M}_n^{-1}$.

\section{Rényi relative entropy and Ising phase transition}
\label{sec:8v}

The topological order entails two locally indistinguishable yet globally orthogonal logical states, even in the presence of finite noise below the threshold. 
This can be detected by the quantum coherent information or quantum relative entropy (Kullback-Leibler divergence)~\cite{Fan23toriccode}. However, von Neumann entropy is not directly accessible in experiment, and we turn to the second Rényi relative entropy accessible in experiment:
\begin{equation}
D_2(\rho_0||\rho_1) = -\ln \text{tr}(\rho_0 \rho_1) + \ln \text{tr} (\rho_0^2) \ .
\end{equation}
The purity can be measured by a swap measurement jointly performed by Alice and Bob:
\begin{equation}
\text{tr}(\rho_\mu\rho_\nu) = \text{tr}( \rho_\mu \otimes \rho_\nu\cdot \text{SWAP}^{\otimes N}) \ ,
\end{equation}
where $\text{SWAP}=(1+X\otimes X+Y\otimes Y+Z\otimes Z)/2$ is the swap operator. 
The SWAP expectation value can be measured by a Bell measurement i.e., applying a controlled-NOT (CNOT) and then measure the two qubits in $X$ and $Z$ basis, respectively, which gives bit-string for $XX$, $ZZ$, as well as $YY$ interaction, see Fig.~5 in the main text. In this way, we can measure the relative Rényi entropy to signal whether the experimentally prepared surface codes remain under the 2-replica threshold (the Ising threshold, slightly larger than the 1-replica Nishimori threshold)~\cite{Fan23toriccode}. 
In classical numerical simulation, we consider independent bit-flip and phase-flip noises to be compared with the experimental results. 

\subsection{Analytical mapping}

The independent phase-flip and bit-flip errors in Alice and Bob lead to the same consequence as independent readout errors. The readout errors can be pulled back the circuit into only phase-flip in Alice and bit-flip in Bob, see Fig.~5 in the main text. 
In this case, $\ln(\rho_0^2)$ becomes $\ln(\rho_{A_0}\cdot \rho_{B_0})$, where $\rho_{A_0}$ and $\rho_{B_0}$ represent the density matrices for Alice and Bob respectively.
Since Alice and Bob suffer from different kind of noise, the contribution of overlap of $\rho_{A_0}$ and $\rho_{B_0}$ from those error chain $E_A$ and $E_B$ such that $E_A$ is a contractible loop and $E_B$ is a loop. In this case, we have: 
\begin{equation*}
   \ln{\rm tr}(\rho_{A_0}\cdot \rho_{B_0}) = P_{E_A}(P^{\prime}_{E_B}+P^{\prime}_{E_BZ_\mathrm{L}})
\end{equation*}
where $P_{E_A}$ and $P_{E_B}$ is given by the partition function of the Ising model:
\begin{align*}
   &P_{E_A} = \sum_{\{s\}} \prod_{<i j>}e^{\beta J_xs_is_j} \ ,\\
   &P^{\prime}_{E_B} = \sum_{\{s\}} \prod_{<i j> }e^{\beta J_zs^{\prime}_is^{\prime}_j} \ .\\
\end{align*}And as the same, for $\ln(\rho_{A_0}\cdot \rho_{B_1})$, we have:
\begin{equation*}
   \ln{\rm tr}(\rho_{A_0}\cdot \rho_{B_1}) = P_{E_A}(P^{\prime}_{E_BX_\mathrm{L}}+P^{\prime}_{E_BX_\mathrm{L}Z_\mathrm{L}}) \ .
\end{equation*}So that finally we have the Rényi relative entropy:
\begin{equation}
   \label{eq:17}
   D = -\ln \frac{{\rm tr}(\rho_{A_0}\cdot \rho_{B_1})}{{\rm tr}(\rho_{A_0}\cdot \rho_{B_0})} = -\ln \frac{P_{E_A}(P^{\prime}_{E_BX_\mathrm{L}}+P^{\prime}_{E_BX_\mathrm{L}Z_\mathrm{L}})}{P_{E_A}(P^{\prime}_{E_B}+P^{\prime}_{E_BZ_\mathrm{L}})}= -\ln \frac{P^{\prime}_{E_BX_\mathrm{L}}+P^{\prime}_{E_BX_\mathrm{L}Z_\mathrm{L}}}{P^{\prime}_{E_B}+P^{\prime}_{E_BZ_\mathrm{L}}} \ .
\end{equation}Note that from the equality: $P^{\prime}_{E_B}P^{\prime}_{E_BX_\mathrm{L}Z_\mathrm{L}} = P^{\prime}_{E_BX_\mathrm{L}}P^{\prime}_{E_BZ_\mathrm{L}}$ the formula ~\ref{eq:17} can be reformulated as:
\begin{equation}
   D = -\ln \frac{P^{\prime}_{E_BX_\mathrm{L}}}{P^{\prime}_{E_B}} = -\ln \frac{\mathcal{\tilde{Z}_I}(J_z)}{\mathcal{Z_I}(J_z)} \ ,
\end{equation}where $\mathcal{Z}_I$ denote the partition function of the Ising model.
The numerical result is shown in Fig.~5 of the main text.

\subsection{Classical error injection for experimental bitstrings}

To investigate the performance of the {\Renyi} relative entropy under different noise levels, we implement a classical post-processing method to simulate increased noise. First, we characterize the intrinsic noise of the system by determining an effective uniform bit-flip error strength $p_0$. This parameter is inferred from the measured purity $\operatorname{tr}\left(\rho_0^2\right)$, using a relationship pre-calibrated through numerical simulations.
Subsequently, we increase the noise level by injecting additional errors into the measured bitstrings. This is achieved by applying a classical bit-flip channel, where each bit is flipped with a probability $p_{\mathrm{flip}}$. The combination of the intrinsic system noise and the injected classical noise results in a boosted effective bit-flip error strength $p$, given by:
\begin{equation}
    p = \left(1-p_0\right)p_{\mathrm{flip}}+(1-p_{\mathrm{flip}})p_0
\end{equation}
By varying the flipping probability $p_{\mathrm{flip}}$ in the range $\left[0, 1/2\right]$, we can scan the effective bit-flip error strength $p$ from its baseline value of $p_0$ up to the theoretical maximum of $1/2$. To ensure robust statistical estimates, this error injection procedure is repeated 100 times for each measured bitstring.

\section{Tunable teleportation phase transition}
\label{sec:effentanglementparameter}

In the classical simulation, the core of the task is to sample the data bitstring $\sigma$, which is given by the Born's rule:
\begin{equation}
P(\sigma) = \langle \bra{\sigma} \mathcal{N} \ket{0_\mathrm{L}}\rangle \ ,
\end{equation}
cast in the double Hilbert space for brevity. When $\mathcal{N} = (1-p_X - p_Y-p_Z) + p_X X^{\otimes 2} + p_Y Y^{\otimes 2} + p_Z Z^{\otimes 2} $ becomes {\it diagonal} Pauli noise, the sampling is straightforward and can be done by an uncorrelated sampling. When $\mathcal{N} = 1-p + p (\frac{X+Z}{\sqrt{2}})^{\otimes 2}$ involves the off-diagonal Pauli noise (sometimes referred to as coherent noise, albeit slightly different from the unitary coherent error), the sampling is less straightforward, but we have mature experience to use a double layer tensor network calculation for the Born sampling. Then we feed the data bitstring to the tensor network decoder to obtain the result.

\subsection{Teleportation in $X$ basis}

Here we take $\theta=\pi/2$. The post-teleportation state that Bob receives (after local correction with Alice's outcomes) effectively suffers from the following noise:
\begin{equation}
\mathcal{N}_Z \circ \mathcal{M}_X \circ \mathcal{N}_X^{\prime}\  (\ket{0_\mathrm{L}}\rangle) \ ,
\end{equation}
where $\mathcal{N}_X^{\prime}=(1-p)+pX^{\otimes2}$ denotes the bit-flip channel arising from the incoherent error $p$ in the preparation of the surface code, $\mathcal{M}_X$ expresses the measurement channel induced by the non-maximally entangled teleportation, and $\mathcal{N}_Z$ denotes the dephasing channel owing to the measurement error of the Alice's qubits. 
We measure the qubits in the $Z$ basis and feed the data bitstring to the classical computer to generate the decoded logical $\langle Z_\mathrm{L}\rangle$. The last $\mathcal{N}_Z$ is absorbed by the measurement basis, and thus the noise action reduces to 
\begin{equation}
\mathcal{N}_X \circ \mathcal{N}_X^{\prime} 
= \left(\cos^2(t)+\sin^2(t)X^{\otimes2}\right) \circ \left((1-p)+pX^{\otimes2}\right)=\cos^2(t_\mathrm{eff})+\sin^2(t_\mathrm{eff})X^{\otimes2}\ ,
\end{equation}
where $\mathcal{N}_X = \cos^2(t)+\sin^2(t)X^{\otimes2}$ is the bit-flip channel caused by the "coherent error" $t$. When only $Z$ measurement is performed, $m$ syndrome is extracted from the data qubits. In this case, 
\begin{equation}
\sin^2(t_\mathrm{eff})=(1-\sin^2(t))p+\sin^2(t)(1-p)=\sin^2(t)+\cos(2t)p \ ,
\end{equation} 
is the effective ``coherent error" rate that creates the $m$ syndrome. Thus we can predict how the logical error $p_{X_\mathrm{L}}$ depends on $p$ and $t$. This effective reparametrization of coherent error is used in Fig.~3a in the main text.

\subsection{Teleportation in a self-dual basis}
The self-duality can maximally enhance the threshold of teleportation. Here we take $\theta=\pi/4$, which results in the following state:
\begin{equation}
\mathcal{N}_{\frac{X-Z}{\sqrt{2}}} \circ \mathcal{M}_{\frac{X+Z}{\sqrt{2}}} \circ \mathcal{N}_X^{\prime}\  (\ket{0_\mathrm{L}}\rangle) \ .
\end{equation}
Here we take the same strategy for experimental measurement and decoding, to check the dependence of the infinite threshold on the noise probability.

If the measurement and local feedback is replaced by a unitary gate, and Alice is traced out afterwards, Bob's surface code can be approximated by the following channel: 
\begin{equation}
   \begin{split}
   \mathcal{N}=&\mathcal{N}_{\frac{X+Z}{\sqrt{2}}} \circ \mathcal{N}_X^{\prime}  \\
   =& \left( \cos^2(t) + \sin^2(t) \left(\frac{X+Z}{\sqrt{2}}\right)^{\otimes 2} \right) \circ \left( (1-p) + pX^{\otimes 2} \right) \\
   =& 
   \left(\frac{\sin^2(t)}{2} + p - \frac{3}{2}p\sin^2(t)\right) X^{\otimes 2} + \frac{p\sin^2(t)}{2} Y^{\otimes 2} + \frac{(1-p)\sin^2(t)}{2} Z^{\otimes 2} +\cdots
   \end{split}
\end{equation}
When the state is further dephased (due to projective measurement in $Z$ basis) such that $Z^{\otimes 2} \to 1$ is applied on the data qubits, the off-diagonal terms that anti-commute with $Z^{\otimes 2}$ are forced to vanish, and the noise reduces to 
\begin{equation}
  \mathcal{N} \to \left(\frac{\sin^2(t)}{2} + p\cos^2(t)\right) X^{\otimes 2} + \cdots \ ,
\end{equation}
therefore the effective coherent error becomes: 
\begin{equation}
   \frac{\sin^2(t_\mathrm{eff})}{2}=\frac{\sin^2(t)}{2} + p\cos^2(t) \Rightarrow
   \sin^2(t_\mathrm{eff}) = \sin^2(t)+p\left(1+\cos(2t)\right) \ ,
\end{equation}
which is used in Fig.~3b in the main text. 

\subsection{Learnability and teleportation transition}
\label{sec:Alicelearn}

Here we compare the learnability transition with the teleportation transition from decoded experimental data. In addition, we also show the data for odd code distances $d=3,5,7$ here, in comparison with the even code distances (mainly discussed in the main text). 
In the teleportation protocol, Alice also does the measurements and gets her bitstrings from which she can in some sense learn the logical state information of the original surface code. Comparing to the teleportable transition from Bob's side, by tuning the entangling strength $t$, Alice also suffers a learnable transition, which can be captured by the entropy of the mixed logical state conditioned upon her measurement outcome. 
\begin{figure}[!htbp]
   \centering
   \includegraphics[width=\linewidth]{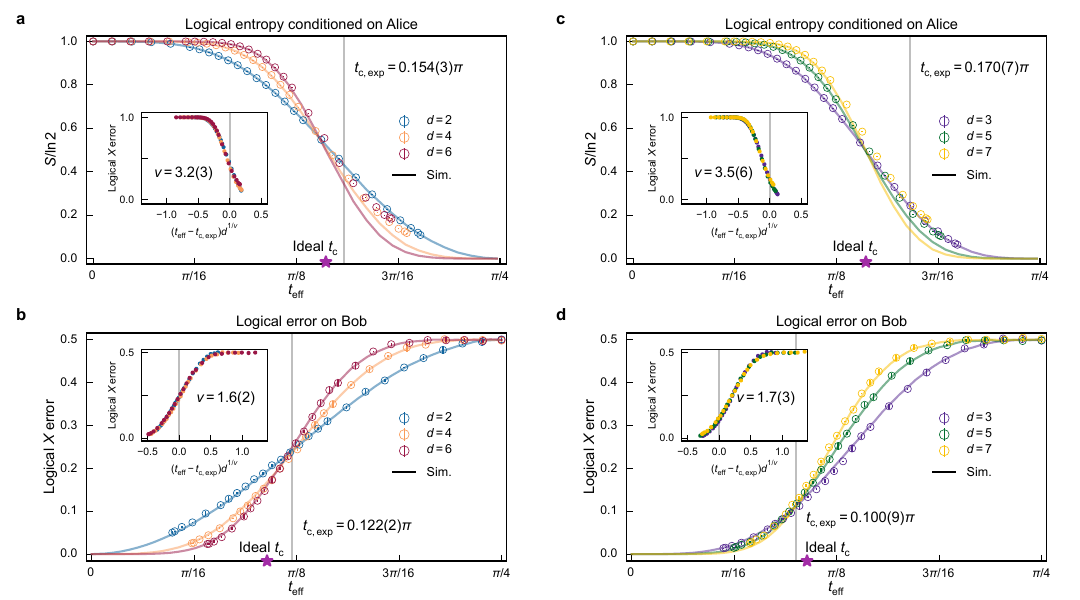}
   \caption{
   {\bf Alice's entropy vs Bob's logical error rate.} The transition point for Alice is $0.143\pi$ while for Bob is $0.107\pi$, they are dual to each other such that $0.143\pi+0.107\pi=0.25\pi$. The mismatch between the Alice experiment data and numerical data is because entropy $S$ is a nonlinear physical quantity and it is sensitive to the correlated errors in the experiment.
   }
   \label{fig:fig_Alice_learning_vs_Bob_pi_over_2}
\end{figure}

\section{Surface code state preparation}
\label{sec:experiment_circuit}

\subsection{Circuit for preparing Alice in a logical state}

\begin{figure*}[htbp]
\centering
\includegraphics[width=\linewidth]{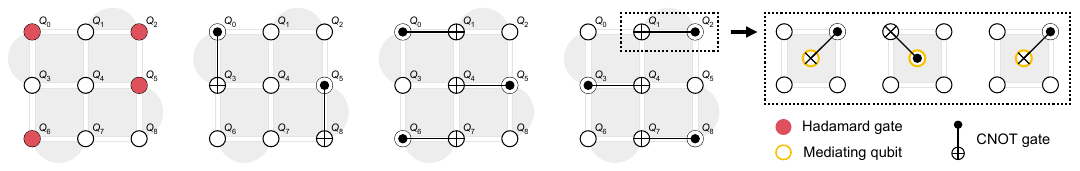}
\caption{
{\bf Circuit schematic for preparing a surface code in the logical state $\ket{0_\mathrm{L}}$.}
All physical qubits are initialized in the product state $\ket{0\cdots0}$. Next, Hadamard gates (red) are applied to representative qubits, which convert the corresponding single-qubit Pauli $Z$ operators in the stabilizer group to single-qubit Pauli $X$ operators. Subsequently, these operators are expanded into the required multi-qubit $X$-type and $Z$-type stabilizers through three layers of CNOT gates. The resulting state, in the absence of error, is the logical state $\ket{0_\mathrm{L}}$. Because Alice's qubits are interleaved with Bob's, these CNOT gates are effectively implemented via mediating qubits (orange circles), as illustrated by the circuit in the dashed box.
}
\label{fig:figS_surface_code_logical_state_preparation}
\end{figure*}

The quantum circuit shown in Fig.~\ref{fig:figS_surface_code_logical_state_preparation} is designed to encode a set of physical qubits into the surface code logical state $\ket{0_\mathrm{L}}$. The process begins with the physical qubits initialized in the product state $\ket{0\cdots0}$. This initial state is stabilized by the set of single-qubit Pauli operators $\{Z_0, Z_1, Z_2, Z_3, Z_4, Z_5, Z_6, Z_7, Z_8\}$. Consequently, it is already an eigenstate of the logical $Z_\mathrm{L}$ operator and all $Z$-type stabilizers ($A_s$) with an eigenvalue of $+1$. However, it is not an eigenstate of the $X$-type stabilizers ($B_f$). The purpose of the encoding circuit is to evolve this initial state into a simultaneous eigenstate of all stabilizers with an eigenvalue of $+1$, including the $X$-type ones. This is accomplished by first applying Hadamard gates to a set of representative qubits, one for each $X$-type stabilizer $B_f$. A Hadamard gate transforms the corresponding $Z_j$ operators into $X_j$ operators. For example, if qubits $Q_0$, $Q_2$, $Q_5$, and $Q_6$ are chosen as representatives, the stabilizers become $\{X_0, Z_1, X_2, Z_3, Z_4, X_5, X_6, Z_7, Z_8\}$. Next, a sequence of CNOT gates extends these single-qubit $X$ operators into the desired multi-qubit $X$-type stabilizers. This leverages the CNOT gate's operator propagation properties: a CNOT from a control qubit $Q_c$ to a target qubit $Q_t$ transforms an operator $X_c$ into $X_c X_t$, and it also transforms an operator $Z_t$ into $Z_c Z_t$. By carefully scheduling the CNOT gates in layers, the stabilizer generators are progressively constructed. The evolution of the generator set after each CNOT layer in Fig.~\ref{fig:figS_surface_code_logical_state_preparation} is as follows:
\begin{align}
    &\{X_0X_3, Z_1, X_2, Z_0Z_3, Z_4, X_5X_8, X_6, Z_7, Z_5Z_8\},\\
    &\{X_0X_1X_3, Z_0Z_1, X_2, Z_0Z_3, Z_4Z_5, X_4X_5X_8, X_6X_7, Z_6Z_7, Z_5Z_8\},\\
    &\{X_0X_1X_3X_4, Z_0Z_1Z_2, X_1X_2, Z_0Z_3, Z_3Z_4Z_5, X_4X_5X_7X_8, X_6X_7, Z_6Z_7Z_8, Z_5Z_8\}.
\end{align}
The final set of generators is equivalent to the stabilizer group defining the logical state $\ket{0_\mathrm{L}}$ of the distance-3 surface code, which is
\[
\{X_0X_1X_3X_4, X_1X_2, X_4X_5X_7X_8, X_6X_7, Z_0Z_3, Z_1Z_2Z_4Z_5, Z_3Z_4Z_6Z_7, Z_5Z_8, Z_6Z_7Z_8\}
\]
The representative qubits and CNOT sequence are chosen to enable parallel gate implementation and minimize the total circuit depth, where the number of the two-qubit gate layers scales as $\left\lfloor \left(d+3\right)/2\right\rfloor$~\cite{Satzinger2021Science}.

However, because Alice's qubits are interleaved with Bob's qubits in order to implement teleportation between two surface codes, the stabilizers cannot be extended simply by applying two-qubit CNOT gates between adjacent qubits belonging solely to Alice. Instead, Bob's qubits must be used to mediate the propagation. Intuitively, this can be accomplished by a sequence of three layers of CNOT gates (Fig.~\ref{fig:figS_surface_code_logical_state_preparation}).
Consider three relevant stabilizers,
\begin{equation}
    \{\cdots X_c I_m I_t \cdots,\; I_c Z_m I_t,\; \cdots I_c I_m Z_t \cdots\},
\end{equation}
which evolve under these three CNOT layers as
\begin{align}
    &\{\cdots X_c X_m I_t \cdots,\; Z_c Z_m I_t,\; \cdots I_c I_m Z_t \cdots\}, \\
    &\{\cdots X_c X_m X_t \cdots,\; Z_c Z_m I_t,\; \cdots I_c Z_m Z_t \cdots\}, \\
    &\{\cdots X_c I_m X_t \cdots,\; I_c Z_m I_t,\; \cdots Z_c Z_m Z_t \cdots\}.
\end{align}
The final stabilizers can be rewritten as
\begin{equation}
    \{\cdots X_c I_m X_t \cdots,\; I_c Z_m I_t,\; \cdots Z_c I_m Z_t \cdots\}.
\end{equation}
Comparing the initial and final sets reveals two key findings. First, the mediating qubit $Q_m$ is returned to its original state $\ket{0}$, which is identified by the persistence of the $I_c Z_m I_t$ stabilizer. Second, the overall effect is identical to that of a single CNOT gate with $Q_c$ as the control and $Q_t$ as the target. Consequently, the number of two-qubit gate layers still scales linearly as $3\left\lfloor \left(d+3\right)/2\right\rfloor$, which can be further reduced by reorganizing the CNOT gates to maximize their parallel implementation.

\subsection{Circuit for preparing both Alice and Bob in a logical state}

When evaluating the relative entropy, both Alice and Bob are prepared in a logical state. Similar to the procedure for encoding only Alice's qubits, all physical qubits are first initialized in the product state $\ket{0\cdots0}$. We then apply Hadamard gates to an appropriate subset of qubits so as to map a corresponding subset of stabilizers from the single-qubit Pauli $Z$ operator to single-qubit Pauli $X$ operator. The central step is then the systematic extension of these single-qubit Pauli $X$ stabilizers to the nonlocal multi-qubit $X$-type stabilizers defining a surface code. 

\begin{figure*}[htbp]
\centering
\includegraphics[width=\linewidth]{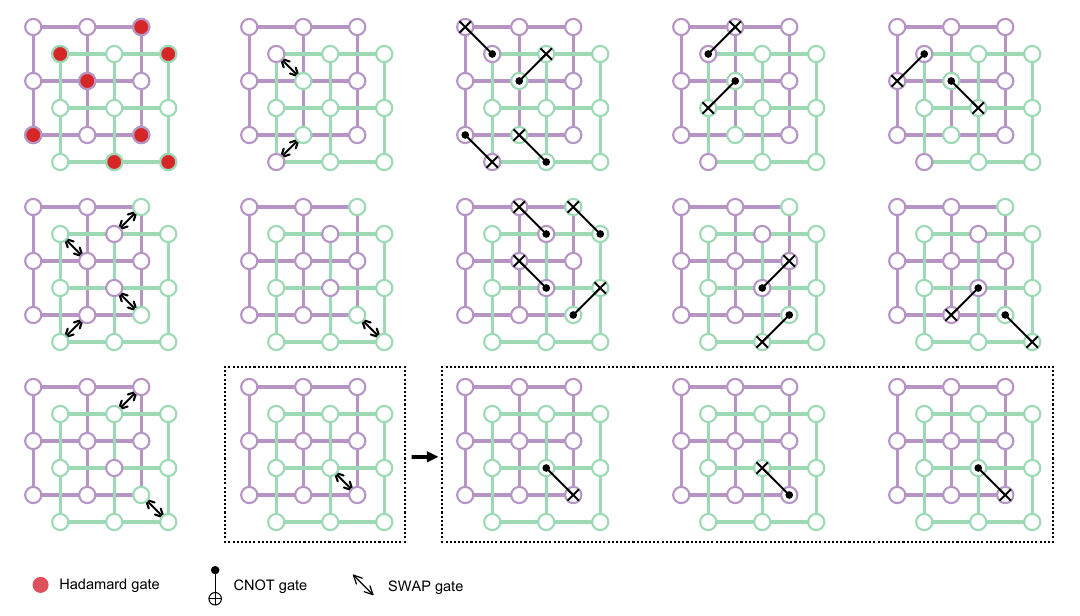}
\caption{
{\bf Circuit schematic for preparing two surface codes in the logical product state $\ket{0_\mathrm{L}} \otimes \ket{0_\mathrm{L}}$.}
Starting from the product state $\ket{0\cdots 0}$ for all physical qubits, Hadamard gates (red circles) are applied to the representative qubits selected from Alice's lattice (purple) and Bob's lattice (green). Subsequently, SWAP gates are employed to temporarily exchange qubits between Alice's and Bob's lattices, thereby bringing the relevant qubits into spatial proximity to enable the application of local CNOT gates. Through successive layers of these CNOT operations, single-qubit Pauli $X$ or $Z$ operators are progressively extended into the required multi-qubit $X$- or $Z$-type stabilizers. Following this, the exchanged qubits are returned to their original lattices via additional SWAP operations. The above SWAP-CNOT-SWAP sequence is repeated iteratively from left to right across the lattice, facilitating the propagation of entanglement throughout the system. This systematic process ultimately prepares Alice and Bob in the logical product state $\ket{0_\mathrm{L}} \otimes \ket{0_\mathrm{L}}$. A SWAP operation is implemented using three layers of CNOT gates, as illustrated in the circuit diagram within the dashed box.
}
\label{fig:figS_surface_code_logical_state_preparation_alice_and_bob}
\end{figure*}

For concreteness, let us focus on the case of code distance of $d=3$ (Fig.~\ref{fig:figS_surface_code_logical_state_preparation_alice_and_bob}). We choose to propagate the support of the $X$-type stabilizers from left to right across the lattice. For larger distances, the number of required two-qubit gates can in principle be reduced by changing the direction and pattern of propagation, for instance by extending the stabilizers from 
the central region toward the boundaries. Here, using three layers of CNOT gates to effectively achieve a single CNOT gate between distant qubits in either Alice's or Bob's lattice proves infeasible. Instead, we introduce SWAP gates to dynamically transfer qubits between the two lattices.

The protocol proceeds as follows. First, two pairs of qubits---one from Alice's lattice and one from Bob's---are swapped. This brings the necessary qubits into proximity, allowing subsequent CNOT gates to extend the initial single-qubit $X$ operators into two- and four-qubit $X$-type stabilizers. We note a circuit optimization here: the initial SWAP gates can be omitted by applying the initial Hadamard gates directly to the target qubits in their swapped positions. 
Subsequently, two swapped qubits are returned to their original lattices via additional SWAP operations. A new set of two qubit pairs is then swapped between Alice's and Bob's system to facilitate the next layers of CNOT gates.
Finally, all qubits are restored to their respective lattices. At this stage, both Alice and Bob are prepared in the logical state $\ket{0_\mathrm{L}}$. 

\subsection{Preparing logical states by measuring stabilizers}
\label{sec:measurementprep}

A more efficient state-preparation protocol leverages stabilizer measurements~\cite{Li2015NewJ.Phys.,Ye2023Phys.Rev.Lett.}, requiring only a constant-depth circuit. In this approach, the data qubits are first initialized in a specific product state, after which a single round of the surface-code cycle is applied to project the data qubits into the logical subspace. 
The resulting logical states may differ from those obtained via a unitary encoding circuit, as the stabilizers are randomly projected to either $+1$ (corresponding to the state $\ket{0}$) or $-1$ (corresponding to the state $\ket{1}$). Nevertheless, these states are logically equivalent up to a Pauli-frame update and therefore give the same logical expectation values in the absence of errors.

\begin{figure}[!htbp]
\centering
\includegraphics[width=3.5in]{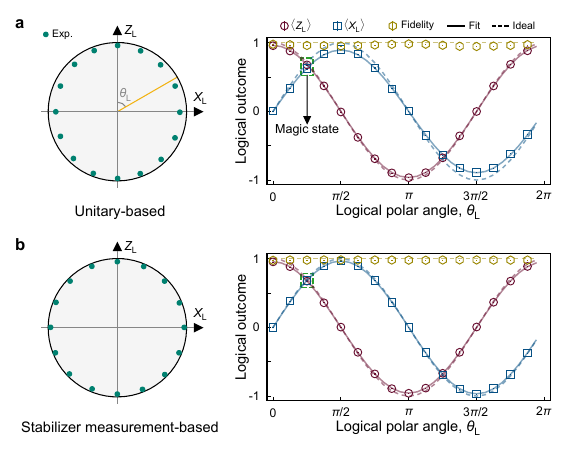}
\caption{
{\bf Comparison of two logical state preparation methods on Alice ($d=3$) with the MWPM decoder.}
{\bf a,} Logical measurement results for the unitary-based state preparation method. The logical outcomes are derived from the same bitstrings used in Fig.~4a, but are decoded here using the MWPM decoder instead of the ML decoder. The reconstructed states projected onto the $XZ$-plane exhibit an elliptical distortion, reflecting an asymmetry in the logical error rates. The average state fidelity is $0.964(11)$.
{\bf b,} Logical measurement results for the stabilizer measurement-based state preparation method ($d = 3$). In contrast to the unitary method, the reconstructed states on the $XZ$-plane display a more circular distribution, reflecting a significant alleviation of error asymmetry. The average state fidelity is $0.981(2)$, demonstrating an improvement over the unitary-based approach.
}
\label{fig:figS_arbitrary_logical_state_Alice_MWPM}
\end{figure}

We implement this approach on our quantum processor, and then validate the generated states using the same method as in Fig.~4 of the main text, see Fig.~\ref{fig:figS_arbitrary_logical_state_Alice_MWPM}. Here, we adopt the minimum-weight perfect matching (MWPM) decoder to generate logical outcomes. The experimental results demonstrate slightly lower logical error rates for all states, with the error asymmetry significantly alleviated. The fitted logical errors are $p_X = 0.022(1)$ and $p_Z = 0.017(1)$, compared to $p_X = 0.019(1)$ and $p_Z = 0.053(1)$ which are extracted from the experimental bitstrings of Fig.~4a in the main text. We note that this protocol is compatible with our experiment when fast mid-circuit measurements and an efficient reset protocol are available, allowing the syndrome qubits to be reused by Bob in subsequent teleportation processes.

\section{Decoding methods}
\label{sec:decoders}

Decoding the logical information from a noisy quantum code is generally an inference problem which allows various decoders, which may or may not yield the optimal threshold. While we have mainly showed the data from tensor network (TN)-based maximum likelihood (ML) decoder in the main text, here we also show the results from two alternative decoders: MWPM decoder and TN-based Bayesian decoder. 

\begin{figure*}[!htbp] 
   \centering
   \includegraphics[width=\textwidth]{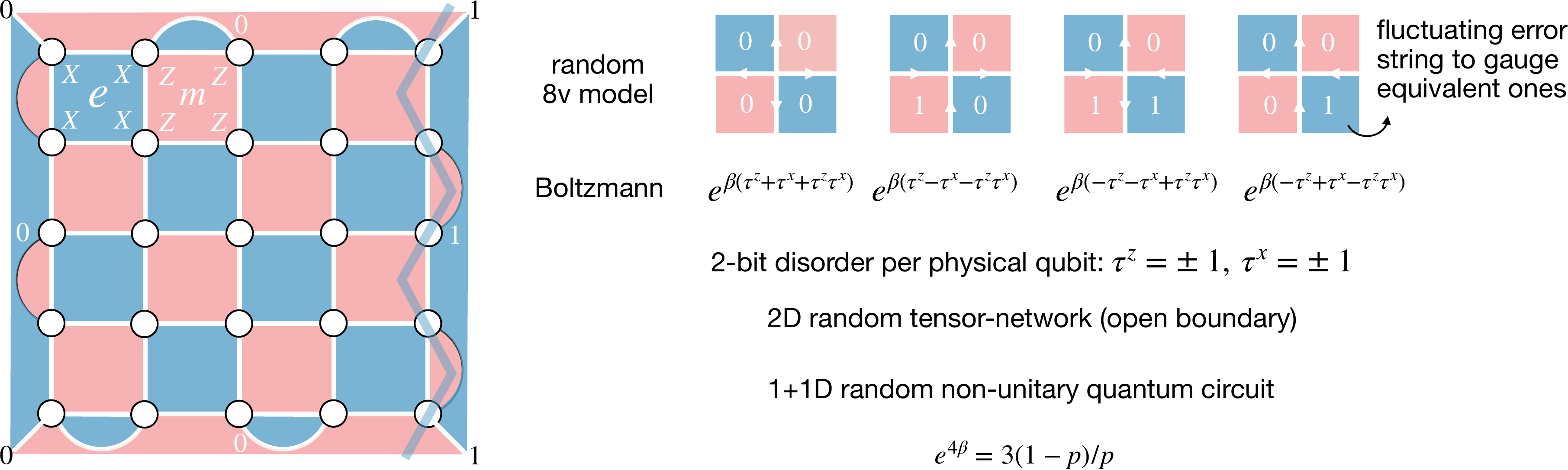} 
   \caption{
   {\bf Random 8-vertex model describing the statistical ensemble of depolarizing error strings hitting the surface code}, taking a 2-step procedure: noisy quantum state $\to$ bilayer Ising model on faces $\to$ 8-vertex model. 
   The two logical operator strings are equivalent to inserting the non-contractible domain walls for the blue and pink Ising layers, respectively. The non-contractible domain walls can be moved to the boundary, reducing to simply the boundary variable being $1$ (otherwise we fix the boundary variable to be $0$ since there is no stabilizer here that can fluctuate the logical error string).
   Thus the bilayer Ising model is placed in fixed boundary condition. 
   Using the standard map from the bilayer Ising model to the vertex model, the face centerred Ising variable is dual to the arrows on the bond, where each vertex expresses a weight given the state of the arrows. One can view the 8-vertex network as a 2D tensor network or a 1+1D quantum circuit, where the arrow direction is viewed as a qubit. The boundary tensor needs some care since the Ising variable is fixed. 
   When serving as a decoder for the experiment, the experimental data bitstring can be input as $\tau$ (up to a gauge transformation) that creates the syndromes. The depolarizing error here can be easily generalized to anisotropic but still diagonal Pauli errors with independent $p_X, p_Y, p_Z$. 
   }
   \label{fig:rand8v}
\end{figure*}

MWPM decoder is well-known to find the shortest path to pair up the syndromes which can be efficiently found by algorithm like Blossom algorithm, and here we use the PyMatching package~\cite{Higgott2025sparseblossom}. It was shown~\cite{Preskill2002} that it yields a threshold $0.103$ as the zero temperature transition point of the random bond Ising model (RBIM), which is extremely close to the optimal threshold $0.109$ as the finite temperature Nishimori transition point of the RBIM. 
The TN-based Bayesian decoder, on the other hand, outputs a probabilistic correction, that minimizes the ``free energy'' of the statistical model by taking into account the ``entropy'' as a key factor. 
Concretely, it exploits a mapping from the noisy quantum state to a classical statistical mechanics model. As illustrated in Fig.~\ref{fig:rand8v}, the noisy quantum state subject to generic depolarizing errors is mapped to a random 8-vertex model~\cite{Bombin12depolarize}. In this representation, each vertex corresponds to a physical qubit and carries Ising variables $\tau^x=\pm 1$ and $\tau^z=\pm 1$ that are determined by the $X$-basis and $Z$-basis measurements. The Ising variable $s= \pm 1$ and $s^{\prime}=\pm 1$ at the face center captures the stabilizer, whose flip can fluctuate an error string to its gauge equivalent counterpart that shares the same syndrome measurement or quantum number.
Therefore $\tau^x s's'=-1$ together describes an $e$-string and $\tau^zss=-1$ describes an $m$-string. The partition function~\cite{Bombin12depolarize} characterizing gauge equivalent error chains under depolarizing noise is  
\begin{equation}
   \mathcal{Z}= \sum_{s s'} e^{\beta (\tau^x s_k's_l' + \tau^zs_is_j+\tau^x\tau^z s_is_js_k's_l')} \ ,
\end{equation}
where $e^{4\beta}=\frac{3(1-p)}{p}$ with $p$ representing the individual Pauli $X$, $Y$ or $Z$ noise. In the main text, we consider the simple version with only bit-flip noise, in this case, the random 8-vertex model reduce to RBIM~\cite{Preskill2002}:
\begin{equation}
   \mathcal{Z}_{\mathrm{RBIM}} = \sum_{\langle i,j \rangle}e^{\beta^{\prime} \tau^z s_is_j}\ ,
\end{equation}
with $e^{2\beta^{\prime}}=\frac{1-p_X}{p_X}$, $p_X$ represents the bit-flip probability. In practice, a quantum processor gains the data bitstring, which is fed into the tensor network decoder shown in Fig.~\ref{fig:rand8v} as $\tau$ variable, the decoder computing the corresponding partition function and yields the decoded logical information.

\begin{figure}[!htbp]
\centering
\includegraphics[width=\linewidth]{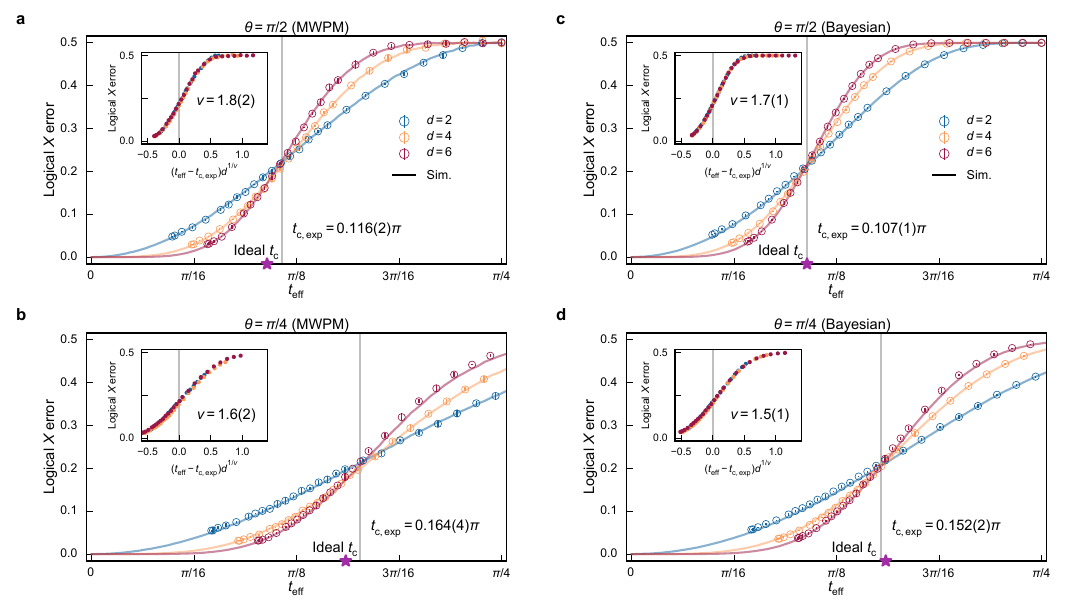}
\caption{
{\bf Results for the teleportation transition using MWPM and Bayesian decoders.}
For each panel, the critical threshold is extracted from a finite-size scaling analysis shown in the inset, and is indicated by the vertical grey solid line.
The stars show the asymptotic thresholds of $t_\mathrm{c}(\theta=\pi/2)=0.1072128(2)\pi$ and $t_\mathrm{c}(\theta=\pi/4)=0.1548014(3)\pi$. 
}
\label{fig:figS_logical_error_TN_and_MWPM_2me_effective_t}
\end{figure}

In Fig.~\ref{fig:figS_logical_error_TN_and_MWPM_2me_effective_t}, we present the decoding results obtained using the MWPM decoder and TN-based Bayesian decoder, applied to the measured bitstrings shown in Fig.~3 of the main text. For the MWPM decoder, the extracted thresholds are $0.116(2)\pi$ at $\theta = \pi/2$ and $0.164(2)\pi$ at $\theta = \pi/4$; these values appear slightly overestimated, which is also observed in the ML decoder. In contrast, the TN-based Bayesian decoder yields thresholds of $0.107(1)\pi$ at $\theta = \pi/2$ and $0.152(2)\pi$ at $\theta = \pi/4$, which agree very closely with the asymptotic theoretical thresholds.

\bibliography{reference}